\documentclass{aa}  
\usepackage{graphicx}
\usepackage{txfonts}
\usepackage{epsfig}
\usepackage{natbib}
\bibpunct{(}{)}{;}{a}{}{,} 

\def \hcm {\hbox {\ifmmode $ cm$^{-2}\else cm$^{-2}$\fi}}

\def\approxgt{\mathrel{\hbox{\rlap{\lower.55ex \hbox {$\sim$}}
        \kern-.3em \raise.4ex \hbox{$>$}}}}
\def\approxlt{\mathrel{\hbox{\rlap{\lower.55ex \hbox {$\sim$}}
        \kern-.3em \raise.4ex \hbox{$<$}}}}

\begin{document}
   \title{The stellar content of low redshift BL Lac host galaxies from multicolour imaging}

   \subtitle{}

   \author{T. Hyv\"onen\inst{1}, J.K. Kotilainen\inst{1}, R. Falomo\inst{2}, E. \"Orndahl\inst{1} 
\and T. Pursimo\inst{3}}

   \offprints{T. Hyv\"onen}

   \institute{Tuorla Observatory, University of Turku, V\"ais\"al\"antie 20, FIN--21500 Piikki\"o, Finland\\
              \email{totahy@utu.fi, jarkot@utu.fi}
         \and
             INAF -- Osservatorio Astronomico di Padova, Vicolo dell'Osservatorio 5, I-35122 Padova, Italy\\
             \email{renato.falomo@oapd.inaf.it}
         \and
             Nordic Optical Telescope, Apartado 474, E-38700 Santa Cruz de La Palma, Santa Cruz de Tenerife, Spain\\
             \email{tpursimo@not.iac.es}
             }

   \date{Received; accepted}

 
  \abstract
   {
We present $B$-band imaging of 18 low redshift (z $\leq$ 0.3) BL Lac objects 
for which their host galaxies were previously resolved in the $R$-band and 
the near-infrared $H$-band. For a subset of the objects, also $U$- and $V$-band imaging is presented.
}
   {
These multiwavelength data are used to investigate the blue--red--near-infrared colours and the colour gradients 
of the host galaxies of BL Lacs in comparison 
with other elliptical galaxies with and without nuclear activity. 
}
   {
For all the BL Lacs observed in the $B$- and $V$-bands, and all objects at $z<0.15$ in the $U$-band, 
the host galaxy is clearly resolved. In all cases galaxies are well represented by an elliptical model, with 
average absolute magnitude $M_B = -21.6\pm0.7$ and average scale length $R_{e}$ = 7.6$\pm$3.2 kpc. 
BL Lac host galaxies are therefore luminous (massive) elliptical galaxies in agreement of 
previous studies in other bands.
}
   {
The best-fit $B$-band Kormendy relation of ($\mu$$_e$ = 3.3 log R$_e$ (kpc) + 18.4 mag arcsec$^{-2}$) 
is in reasonable agreement with that obtained for normal ellipticals and radio galaxies. 
This structural and dynamical similarity indicates that all massive elliptical galaxies can experience 
nuclear activity without significant perturbation of their global structure. 
The distributions of the integrated blue/near--infrared colour (with average $B$--$H$ = 3.5$\pm$0.5) and colour gradient 
(with average $\Delta$($B$--$R$)/$\Delta$(log r) = -0.14$\pm$0.75) of the BL Lac hosts 
are much wider than those for normal ellipticals, and most BL Lac objects have 
bluer hosts and/or steeper colour gradients than those in normal ellipticals. 
}
   {
The blue colours are likely caused by a young stellar population component, 
and indicates a link between star formation caused by an 
interaction/merging event and the onset of the nuclear activity. 
This result is corroborated by stellar population modelling, indicating a presence of young/intermediate age 
populations in the majority of the sample, in agreement with low redshift quasar hosts. 
The lack of strong signs of interaction may require 
a significant time delay between the event with associated star formation episodes and 
the start of the nuclear activity.
}


   \titlerunning{The stellar content of low redshift BL Lac host galaxies}
   \authorrunning{Hyv\"onen et al.}

   \maketitle

\section{Introduction}

BL Lac objects are an extreme class of active galactic nuclei (AGN) characterized 
by luminous, variable and polarized continuum emission across the 
electromagnetic spectrum and strong core-dominated radio emission with apparent 
superluminal motion \citep[e.g.,][]{koll92}. These properties suggest that they are 
strongly beamed objects dominated by synchrotron emission from a relativistic jet 
aligned close to the line of sight \citep{blan78}. BL Lac objects have many similarities to 
flat spectrum radio quasars (FSRQ) and they are often grouped together as blazars. 
According to the unified model of radio-loud AGN \citep{urry95}, the parent population 
of BL Lac objects and FSRQs are low luminosity core-dominated FR I radio galaxies (RG), 
and high luminosity lobe-dominated FR II RGs, respectively. 
Consequently, their orientation independent properties, such as host galaxies 
and environments, should be identical to those of their parent populations.

A number of optical \citep[e.g.,][]{falo99,urry00,nils03,heid04} and 
near-infrared (NIR) \citep[e.g.,][]{koti98b,scar00b,cheu03,koti04,koti05} imaging studies have 
shown that virtually all nearby $(z<0.5)$ BL Lac objects are hosted in large and luminous elliptical galaxies, 
with average $M_R \sim-24.0$ and average scale length $R_{e}\sim10$ kpc, similarly to 
both FR I and FR II RGs \citep{govo00}. BL Lac hosts are much brighter than $L^{*}$ 
galaxies (the characteristic luminosity of the Schechter luminosity function: 
$M^{*}_{R}=-21.2$, \citet{gard97,naka03}). 
While the morphology, such as jets and close companions, of some of the hosts indicates a recent interaction, 
the large majority of them are indistinguishable from inactive massive ellipticals at similar redshift \citep{scar00b}.

Until recently, however, imaging of BL Lac host galaxies was obtained in one band only 
(usually $R$-band). Therefore, little colour information exist for them, 
especially in the blue domain of the spectrum, where only a few objects have been studied in 
the $B$-band and even fewer in the $U$-band. \citet{koti04} observed 
a sample of 23 low redshift ($z<0.3$) BL Lac objects in the $H$-band and combined with 
previous $H$-band imaging \citep{koti98b,scar00b,cheu03} and optical $R$-band data from literature 
\citep{falo99,urry00}, investigated the 
integrated $R$--$H$ colours and colour gradients of a sample of 41 BL Lac host galaxies. 
They found that BL Lac host galaxies appear to be systematically bluer than inactive ellipticals 
and have a much wider 
distribution of host galaxy $R$--$H$ colour (average $R$--$H=2.2\pm0.4$) and steeper colour 
gradient (average $\Delta(R$--$H)/\Delta(logr)=-0.38\pm0.60$) than those for inactive elliptical 
galaxies with dominant old stellar population \citep{pele90}. 
Similar behaviour has been found for the colours and colour gradients in low redshift RGs 
\citep{govo00} and in low and intermediate redshift AGN and quasars with elliptical hosts 
\citep{scha00,orn03,jahn04,sanc04}. 
The blue colours are most likely caused by a young stellar population, indicating 
a recent star formation (SF) episode, possibly triggered by interaction or merging 
between galaxies. The wide distribution of colours probably reflects an 
object-to-object difference in the age of the most recent SF episode. 
Note that there is also spectroscopic evidence for young/intermediate age populations in AGN hosts 
\citep[e.g.,][]{nola01,raim05}. 

In this study, we present multicolour optical imaging of a subsample of 18 BL Lac objects from the large, 
homogeneous sample of 41 sources for which high resolution $R$- and $H$-band imaging exists 
\citep{koti98b,scar00b,cheu03,koti04}. 
Most of the observed objects have bluer $R$--$H$ host colour than inactive ellipticals 
and blue band observations are paramount to assess whether their blue colours are caused by 
a young stellar population. 
All the 18 BL Lac objects were observed in the $B$-band, while for a subsample, we obtained also 
$U$- and $V$-band imaging. The $U$-band observations were mainly restricted 
to the most nearby targets in the sample. Combining the deep high spatial resolution 
$U$-, $B$- and $V$-band imaging with the existing $R$- and $H$-band data, 
we are able to derive the colours and colour gradients over an extended wavelength range, 
where the $H$-band is sensitive to old stellar population that dominates the host galaxy mass, 
while the blue part of the spectrum 
traces the contribution from young stellar populations to the excess blue light in the host galaxies. 
We also derive structural properties, such as morphology and effective radius, as well as 
the absolute magnitude of the host galaxies in each observed band. The obtained $UBVRH$ broad band colours 
are used in conjunction with stellar synthesis population models to estimate the ages of 
the most recent SF episode in the host galaxies.

In Section 2, we describe the sample, observations, data reduction, and methods of analysis. 
In Section 3, we present the results and discussion concerning the properties of the host galaxies. 
Summary and conclusions are given in Section 4. Throughout this paper, 
$H_0=70$ kms$^{-1}$Mpc$^{-1}$, $\Omega_{m}=0.3$ and $\Omega_{\Lambda}=0.7$ cosmology is used.

\section{Observations, data reduction, and analysis}

The observations were carried out during several observing runs with different telescopes, 
with most of the observations done at the 2.5m Nordic Optical Telescope (NOT) with the $ALFOSC$ instrument 
using Bessel $U$, $B$ and $V$ broad band filters. 
Nine targets were observed only in the $B$-band, five in the $UBV$-bands and four in the $UB$-bands. 
One target (MRK 421) was additionally observed in the $H$-band, to complement 
the sample presented in \citet{koti04}. 
The B-band image of 3C 371, published in \citet{nils97}, was kindly provided for us by K. Nilsson. 
A summary of the telescopes and instruments used is given in Table~\ref{teles}. 
Seeing during the observations varied between 0.9 and 2.3 arcsec FWHM (average $1.5\pm0.5$ arcsec 
and median 1.3 arcsec). For each target, several integrations (typically of 600s duration) 
were obtained. The journal of the observations and properties of the sample are given in Table~\ref{sample}.

\setlength{\tabcolsep}{0.8mm}
\begin{table}
\caption{Properties of the telescopes and instruments.$^{\mathrm{a}}$}
\label{teles}
\begin{tabular}{llrl}
\hline
\noalign{\smallskip}
Band & Telescope+instrument & $N_{obj}$ & Scale \\
& & & arcsec $px^{-1}$ \\
(1) & (2) & (3) & (4) \\
\noalign{\smallskip}
\hline
\noalign{\smallskip}
$U$,$B$,$V$ & NOT/ALFOSC & 26 & 0.190  \\
$U$,$B$     & NOT/MOSCA  & 3  & 0.217 \\
$H$         & NOT/NOTCam & 1  & 0.235 \\
$U$,$B$     & ESO NTT/EMMI   & 3  & 0.273 \\
\noalign{\smallskip}
\hline
\end{tabular}
\begin{list}{}{}
\item[$^{\mathrm{a}}$] 
Column (1) gives the observed band; (2) the telescope and instrument used; 
(3) the number of observed targets, and (4) the spatial scale of 
the instrument. 
\end{list}
\end{table}
\setlength{\tabcolsep}{1.5mm}

\begin{table*}
\centering
\caption{The sample and the journal of observations.$^{\mathrm{a}}$}
\label{sample}
\begin{tabular}{lllllrlll}
\hline
\noalign{\smallskip}
Name & $z$ & $V$ & $M_B$ & Filter & $T_{exp}$ & Date & FWHM & Photometric?\\
& & & & (sec) & & (arcsec) & \\
(1) & (2) & (3) & (4) & (5) & (6) & (7) & (8) & (9) \\
\noalign{\smallskip}
\hline
\noalign{\smallskip}
1ES 0229+200   & 0.139 & 18.0 & -21.7 & $U$ & 2700 & 12/11/2004 & 1.7 & Y  \\
               &       &      &       & $B$ & 900  & 12/11/2004 & 1.9 & Y  \\
               &       &      &       & $V$ & 1800 & 12/11/2004 & 2.0 & Y  \\
PKS 0521-365   & 0.055 & 14.6 & -22.3 & $B$ & 480  & 14/03/2005 & 1.4 & Y  \\
PKS 0548-322   & 0.069 & 15.5 & -22.0 & $B$ & 900  & 15/03/2005 & 1.3 & Y  \\
1H 0706+591    & 0.125 & 19.5 & -21.0 & $U$ & 1600 & 27/01/2004 & 1.2 & Y  \\
               &       &      &       & $B$ & 1500 & 27/01/2004 & 0.9 & Y  \\
               &       &      &       & $V$ & 600  & 27/01/2004 & 1.2 & Y  \\
MRK 421        & 0.031 & 13.8 & -22.9 & $U$ & 900  & 08/03/2005 & 0.9 & N \\
               &       &      &       & $B$ & 300  & 08/03/2005 & 0.9 & N \\
               &       &      &       & $H$ & 150  & 25/03/2005 & 1.8 & Y  \\
MRK 180        & 0.045 & 15.0 & -22.1 & $U$ & 900  & 06/05/2002 & 1.2 & Y  \\
               &       &      &       & $B$ & 1600 & 10/03/2003 & 1.2 & Y  \\
1ES 1212+078   & 0.136 & 16.1 & -23.6 & $B$ & 1800 & 08/03/2005 & 0.9 & N  \\
MS 1229.2+6430 & 0.164 & 16.9 & -21.7 & $B$ & 1800 & 10/03/2004 & 2.1 & Y  \\
1ES 1255+244   & 0.140 & 15.4 & -24.3 & $B$ & 1800 & 09/03/2005 & 1.9 & Y  \\
PG 1418+546    & 0.152 & 15.7 & -23.7 & $U$ & 900  & 07/05/2002 & 2.1 & Y  \\
               &       &      &       & $B$ & 700  & 07/05/2002 & 1.3 & Y  \\
1ES 1426+428   & 0.129 & 16.5 & -22.5 & $B$ & 1600 & 10/03/2003 & 1.8 & Y  \\
1ES 1440+122   & 0.162 & 17.0 & -22.9 & $B$ & 1800 & 08/03/2005 & 0.9 & N \\
AP LIBRAE      & 0.049 & 14.8 & -21.7 & $U$ & 700  & 07/05/2002 & 2.0 & Y  \\
               &       &      &       & $B$ & 900  & 15/03/2005 & 1.1 & Y  \\
MS 1552.1+2020 & 0.273 & 17.7 & -23.5 & $B$ & 1200 & 09/03/2005 & 1.9 & Y \\
MRK 501        & 0.034 & 13.8 & -22.4 & $U$ & 600  & 07/05/2002 & 1.2 & Y  \\
               &       &      &       & $B$ & 240  & 08/03/2005 & 1.1 & N \\
               &       &      &       & $V$ & 180  & 08/03/2005 & 1.2 & N \\
I Zw 187       & 0.055 & 16.4 & -21.1 & $U$ & 500  & 07/05/2002 & 2.0 & Y  \\
               &       &      &       & $B$ & 900  & 25/09/2003 & 1.1 & Y \\
               &       &      &       & $V$ & 300  & 08/03/2005 & 1.3 & N \\
3C 371         & 0.051 & 14.2 & -22.6 & $U$ & 1200 & 01/10/2000 & 2.1 & Y  \\
	       &       &      &       & $B$ & 3600 & 02/05/1997 & 0.8 & Y  \\
               &       &      &       & $V$ & 480  & 08/03/2005 & 1.3 & N  \\
BL LACERTAE    & 0.069 & 14.7 & -22.4 & $B$ & 720  & 01/10/2000 & 2.3 & Y  \\
\noalign{\smallskip}
\hline
\end{tabular}
\begin{list}{}{}
\item[$^{\mathrm{a}}$]
Column (1) gives the name of the BL Lac object; (2) the redshift; 
(3) the $V$-band apparent magnitude; (4) the $B$-band absolute magnitude; (5) the filter used; 
(6) the total integration time; (7) the date of the observation; (8) the seeing FWHM and 
(9) photometric (Y) or non-photometric (N) conditions.
\end{list}
\end{table*}

Data reduction was performed in standard fashion using IRAF\footnote{IRAF is distributed by 
the National Optical Astronomy Observatories, which are operated by 
the Association of Universities for Research in Astronomy, Inc., 
under cooperative agreement with the National Science Foundation.}. This consisted of 
bias subtraction, flat field division and cosmic ray rejection. 
For each night, bias and flat field images were made from several median combined bias frames 
and twilight flat field exposures and cosmic rays were rejected using IRAF procedures. 
Finally, individual images were aligned and combined to form the final image of the target. 
The observations were mostly done in photometric conditions 
(see Table~\ref{sample})
and several photometric standard stars from \citet{land92} were observed during each night. 
Some targets were, however, observed during non-photometric nights and for these objects additional short 
exposures were subsequently obtained in photometric conditions to calibrate these frames 
using reference stars in the field. 
$K$-correction from \citet{pogg97} was applied to the host galaxy magnitudes, 
but not to the nuclear magnitudes, since the nuclear component can be assumed to have a power-law spectrum 
($f_\nu \propto \nu^{-\alpha}$) with $\alpha\sim-1$. Absolute magnitudes were also corrected for 
interstellar extinction calculated for each band from the $R$-band extinction 
coefficient from \citet{urry00}.

\begin{figure}
\centering
\resizebox{8cm}{13cm}{\includegraphics[bb=1.cm 9.cm 11.5cm 24.7cm,clip]{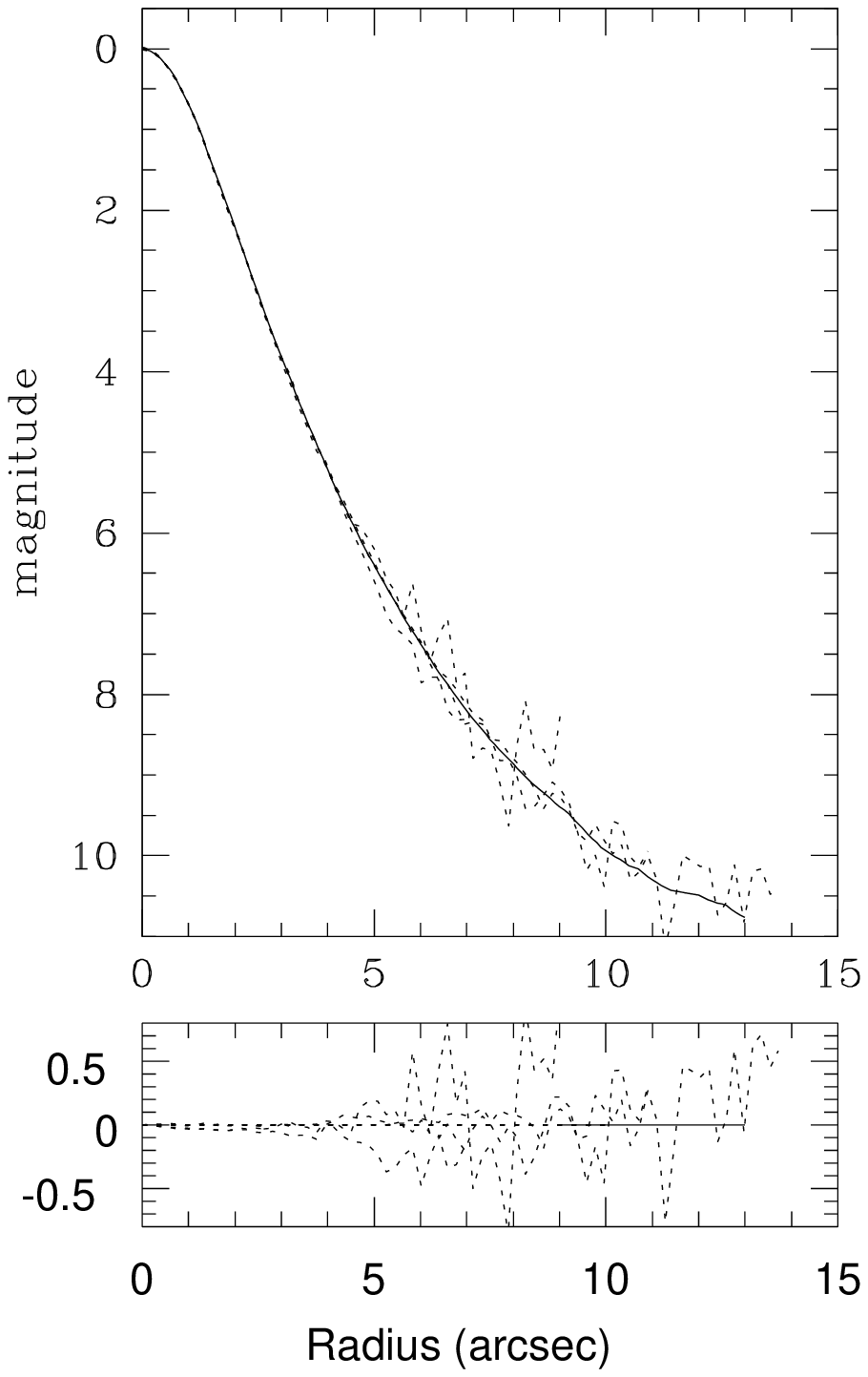}}
\caption[]{Upper panel: 
Comparison of the PSF profile extracted from field stars (the solid line) 
with the profiles of individual stars (dotted lines) in the field of BL Lac object. 
Lower panel: The difference between the profiles of individual stars and the PSF model. 
\label{psf_comp}}
\end{figure}

To derive the properties of the host galaxies, azimuthally averaged 1D radial luminosity profiles 
were extracted for each BL Lac object and for a number of field stars. 
Any obvious extra features, such as nearby companions and/or foreground stars, were masked out from 
the image to avoid contamination of the radial luminosity profile. To obtain an 
accurate model for the nuclear region of the targets, it is important to have a well defined 
point spread function (PSF). 
In most cases there were a number of suitable stars in the relatively large field of view 
surrounding the objects to form a reliable PSF. The core and the wing of 
the PSF were derived from a faint and a bright star in the frame, respectively, and they were combined 
to form the final PSF model. This final PSF was compared with the profiles of individual stars 
in the frame to assure that this procedure resulted in a good and stable representation of the true PSF. 
A representative case of this PSF comparison is shown in Fig.~\ref{psf_comp}, where the individual 
stellar profiles are in good agreement with the adopted PSF model well into the domain 
where the host galaxy becomes dominant over the PSF. The only exception was the 
field of MRK 421 where there are no suitable stars available. In this case, the PSF was estimated 
using standard stars observed during the same night with similar seeing conditions.

The luminosity profiles were decomposed into a point source (represented by the PSF) 
and an elliptical galaxy components by 
an iterative least-squares fit to the observed profile. There are three free parameters in the fit: 
the PSF normalization, the host galaxy normalization and the effective radius of the host galaxy.
The data were fit using the $r^{1/4}$ de Vaucouleurs law for elliptical galaxies to represent the host galaxy. 
The host galaxy was considered to be resolved if the PSF + host galaxy fit resulted 
in a considerably lower $\chi^{2}$ value than the PSF fit only. 
The uncertainty in the derived host galaxy magnitudes was estimated to be $\sim0.2$ mag.


\section{Results}

From the images 1D azimuthally averaged radial luminosity profiles of the BL Lac objects were extracted in 
the $U$-, $B$- and $V$-band. The profiles together with the best-fit model overlaid are presented in 
Appendix in online version of the paper. 

We were able to clearly resolve the host galaxy in all objects 
in the $B$- and $V$-bands, and in 8/9 objects in the $U$-band.  
The only target that remained unresolved in the $U$-band, PG 1418+546, 
is one of the most distant objects in the sample. 
The results of the modelling are summarized in Table~\ref{results}.





\begin{table*}
\centering
\caption{Properties of the host galaxies.$^{\mathrm{a}}$}
\label{results}
\begin{tabular}{lccccccccccc}
\hline
\noalign{\smallskip}
Name&Filter & A & $z$ & $m_{nuc}$ & $m_{host}$ & $\mu_e$ & $r_e$ & $R_e$ & $M_{nuc}$ & $M_{host}$ & $N/H$ \\
& & mag & & mag & mag & mag arcsec$^2$ & arcsec & kpc & mag & mag & \\
(1) & (2) & (3) & (4) & (5) & (6) & (7) & (8) & (9) & (10) & (11) & (12) \\
\noalign{\smallskip}
\hline
\noalign{\smallskip}
1ES 0229+200   & $U$ & 0.84 & 0.139 & 19.22 & 18.83 & 20.94 & 2.75 & 6.55 &-20.6 &-21.5& 0.44\\
               & $B$ & 0.71 &       & 19.58 & 18.57 & 22.39 & 5.65 & 13.4 &-20.2 &-21.9& 0.21\\
               & $V$ & 0.54 &       & 19.67 & 17.58 & 21.72 & 4.90 & 11.7 &-19.9 &-22.2& 0.12\\
PKS 0521-365   & $B$ & 0.25 & 0.055 & 16.58 & 16.17 & 21.50 & 7.60 & 8.02 &-20.6 &-21.3& 0.52\\
PKS 0548-322   & $B$ & 0.17 & 0.069 & 17.42 & 16.11 & 22.31 & 9.40 & 12.2 &-20.2 &-21.8& 0.23\\
1H 0706+591    & $U$ & 0.50 & 0.125 & 17.67 & 18.55 & 21.90 & 3.60 & 7.83 &-21.6 &-21.1& 1.6\\
               & $B$ & 0.42 &       & 18.38 & 17.86 & 21.21 & 3.50 & 7.63 &-20.8 &-22.0& 0.33\\
               & $V$ & 0.32 &       & 18.10 & 16.57 & 20.40 & 3.65 & 7.96 &-21.0 &-22.8& 0.19\\
MRK 421        & $U$ & 0.16 & 0.031 & 13.41 & 15.22 & 21.29 & 7.80 & 4.80 &-22.4 &-20.7& 4.8\\
               & $B$ & 0.14 &       & 14.92 & 15.43 & 21.78 & 8.90 & 5.48 &-20.9 &-20.5& 1.4\\
               & $H$ & 0.00 &       & 15.18 & 14.29 & 16.82 & 6.95 & 4.45 &-20.5 &-24.9& 0.02\\
MRK 180        & $U$ & 0.16 & 0.045 & 14.83 & 15.81 & 20.83 & 4.95 & 4.33 &-21.8 &-21.0& 2.1\\
               & $B$ & 0.14 &       & 16.44 & 15.45 & 21.91 & 9.45 & 8.27 &-20.2 &-21.4& 0.33\\
1ES 1212+078   & $B$ & 0.13 & 0.136 & 19.17 & 18.53 & 22.09 & 3.80 & 8.89 &-19.9 &-21.3& 0.28\\
MS 1229.2+6430 & $B$ & 0.17 & 0.164 & 19.14 & 18.19 & 21.28 & 3.15 & 8.57 &-20.4 &-22.2& 0.19\\
1ES 1255+244   & $B$ & 0.13 & 0.140 & 19.45 & 18.33 & 21.90 & 3.55 & 8.50 &-19.7 &-21.6& 0.17\\
PG 1418+546    & $U$ & 0.09 & 0.152 & 15.95 &       &       &      &      &-23.4 &     &     \\
               & $B$ & 0.08 &       & 17.20 & 18.14 & 20.25 & 1.90 & 4.87 &-22.1 &-21.9& 1.2\\
1ES 1426+428   & $B$ & 0.13 & 0.129 & 17.67 & 17.50 & 19.35 & 1.40 & 3.13 &-21.3 &-22.1& 0.48\\
1ES 1440+122   & $B$ & 0.13 & 0.162 & 18.14 & 19.17 & 22.43 & 3.65 & 9.84 &-21.4 &-21.2& 1.2\\
AP LIBRAE      & $U$ & 0.84 & 0.049 & 15.26 & 16.10 & 19.07 & 2.55 & 2.42 &-22.2 &-21.6& 1.7\\
               & $B$ & 0.71 &       & 16.88 & 16.52 & 19.74 & 2.90 & 2.75 &-20.5 &-21.1& 0.58\\
MS 1552.1+2020 & $B$ & 0.29 & 0.273 & 19.22 & 19.07 & 20.09 & 3.40 & 13.53 &-21.7 &-23.1& 0.28\\
MRK 501        & $U$ & 0.16 & 0.034 & 14.59 & 14.43 & 21.07 & 8.90 & 5.98 &-21.4 &-21.7& 0.76\\
               & $B$ & 0.13 &       & 14.89 & 14.20 & 20.76 & 9.45 & 6.35 &-21.1 &-22.0& 0.44\\
               & $V$ & 0.10 &       & 14.43 & 13.00 & 19.68 & 9.45 & 6.35 &-21.5 &-23.2& 0.21\\
I Zw 187       & $U$ & 0.25 & 0.055 & 16.53 & 17.56 & 21.11 & 2.65 & 2.80 &-20.5 &-19.8& 1.9\\
               & $B$ & 0.21 &       & 17.12 & 17.14 & 21.65 & 4.35 & 4.59 &-20.0 &-20.3& 0.76\\
               & $V$ & 0.16 &       & 16.39 & 15.82 & 20.93 & 5.15 & 5.43 &-20.7 &-21.4& 0.52\\
3C 371         & $U$ & 0.45 & 0.051 & 15.33 & 16.17 & 22.13 & 8.95 & 8.80 &-21.9 &-21.2& 1.9\\
               & $B$ & 0.38 &       & 15.86 & 15.98 & 21.12 & 6.15 & 6.05 &-21.3 &-21.4& 0.91\\
               & $V$ & 0.29 &       & 14.98 & 14.94 & 20.12 & 5.60 & 5.51 &-22.1 &-22.2& 0.91\\
BL LACERTAE    & $B$ & 1.61 & 0.069 & 14.71 & 17.35 & 19.79 & 3.20 & 4.15 &-24.3 &-22.0& 8.32\\
\noalign{\smallskip}
\hline
\end{tabular}
\begin{list}{}{}
\item[$^{\mathrm{a}}$]
Column (1) gives the name of the object; (2) the filter; (3) the interstellar extinction in the $U,B$- and $V$-bands; 
(4) the redshift of the object; (5) the apparent nuclear magnitude; 
(6) the apparent host galaxy magnitude; (7) surface brightness at the effective radius 
(8) and (9) the apparent and absolute effective radius, respectively; (10) the absolute nuclear magnitude; 
(11) the absolute host galaxy magnitude and (12) the absolute nuclear/host luminosity ratio.
\end{list}
\end{table*}
\setlength{\tabcolsep}{1.5mm}

\subsection{Luminosities and sizes of the host galaxies}

\begin{figure}
\centering
\includegraphics[width=15cm]{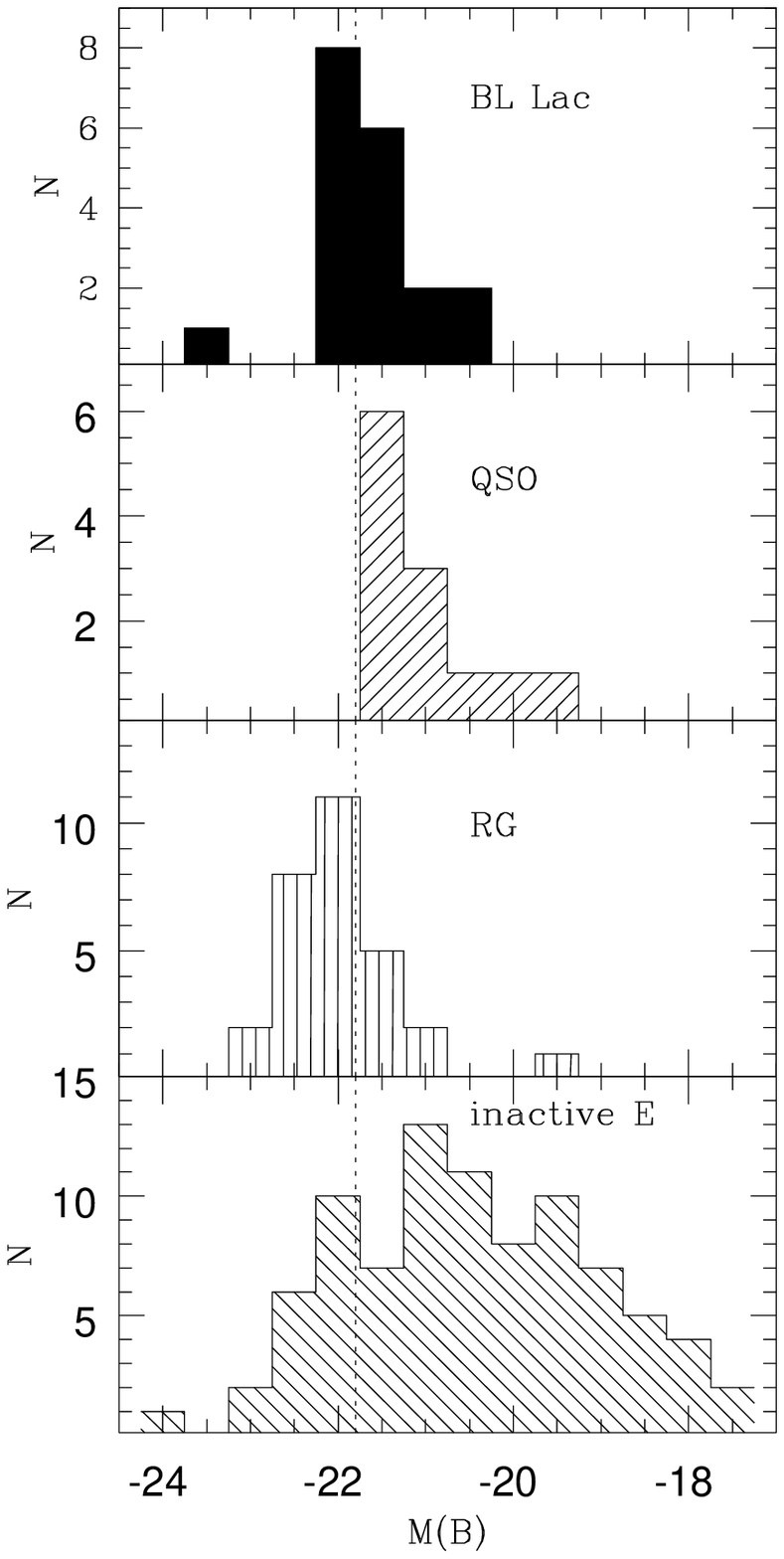}
\caption[]{Histogram of the absolute $B$-band magnitudes of the BL Lac hosts (top panel; this work), 
low redshift ($z<0.2$) RLQ and RQQ hosts (second panel; \citep{jahn04}), low redshift RGs 
(third panel; \citep{govo00}) and inactive elliptical galaxies 
(bottom panel; \citet{pele90,colb01} and \citet{bowe92a}). The vertical short dashed line represent the luminosity of the 
L$^*$ galaxies.
\label{Mbhist}}
\end{figure}

Fig.~\ref{Mbhist} shows the distribution of the $B$-band absolute magnitudes of the BL Lac 
host galaxies (this work), low redshift radio-loud (RLQ) and radio-quiet (RQQ) quasar hosts from 
\citet{jahn04}, low redshift RGs from \citet{govo00} 
and inactive elliptical galaxies from \citet {pele90}, \citet{colb01} and \citet{bowe92a}. 
The average $B$-band absolute magnitudes of the BL Lac objects, quasar and RG host galaxies in these samples are presented in 
Table~\ref{comparison}, where all absolute quantities were transformed into the cosmology adopted here. 

Since the selection criteria of the various samples are somewhat non-homogeneous, there is a 
possibility of selection effects. They are likely to affect absolute magnitudes but since the main 
focus of this study is on the colour properties of BL Lac hosts as compared with other ellipticals, 
our main results are only marginally affected by selection effects. 
Furthermore, our BL Lac sample is well matched to RGs and quasars in redshift 
(see Table~\ref{comparison}) to minimize any bias introduced by possible evolutionary effects.

The average $B$-band absolute magnitude of the 18 BL Lac hosts is $M_B=-21.6\pm0.7$, i.e. $\sim$1 mag 
brighter than for low redshift quasar hosts (average $M_B=-20.9\pm0.4$ \citep{jahn04}). 
All the BL Lac hosts are in the luminosity range between $M^{*}_{B}+0.5$ and $M^{*}_{B}+2.5$, 
where $M^{*}_{B}=-20.8$ \citep{gard97,naka03} is the characteristic luminosity of the Schechter luminosity 
function for elliptical galaxies. 
The average luminosity of the BL Lac hosts is in agreement with inactive isolated early-type galaxies 
($M_B=-21.5\pm0.7$; \citep{colb01}) but is $\sim1$ mag brighter than ellipticals observed 
by \citet{pele90} ($M_B=-20.9\pm1.6$). The distribution of inactive ellipticals has a tail toward 
fainter magnitudes that is not present in the BL Lac sample.


\begin{table*}
\centering
\caption{Comparison of the average $B$-band host galaxy properties.$^{\mathrm{a}}$}
\label{comparison}
\begin{tabular}{lrccc}
\hline
\noalign{\smallskip}
Name & $N$ & $<z>$ & $<M_{B}(host)>$ & $<R_{e}>$ \\
& & & mag & kpc \\
(1) & (2) & (3) & (4) & (5) \\
\noalign{\smallskip}
\hline
\noalign{\smallskip}
BL Lacs (this work)        & 18 & $0.104\pm0.064$ & $-21.6\pm0.7$ & $7.6\pm3.2$  \\
Quasars \citep{jahn04}     & 12 & $0.094\pm0.043$ & $-20.9\pm0.7$ & $6.0\pm2.7$  \\
FR I RGs \citep{govo00}    & 17 & $0.069\pm0.031$ & $-22.1\pm0.6$ & $15.3\pm9.2$    \\
FR II RGs \citep{govo00}   &  6 & $0.050\pm0.016$ & $-21.9\pm0.6$ & $11.4\pm7.7$    \\
\noalign{\smallskip}
\hline
\end{tabular}
\begin{list}{}{}
\item[$^{\mathrm{a}}$]
Column (1) gives the sample; (2) the number of objects; (3) the average redshift; 
(4) the average host galaxy magnitude, and (5) the average effective radius ($<R_e>$ for RGs are in $R$-band). 
\end{list}
\end{table*}
\setlength{\tabcolsep}{1.5mm}

According to the unified model of AGN \citep{urry95}, BL Lac objects are physically similar to FR I RGs but 
viewed from a different orientation. The classification of BL Lac objects with FR Is is based on their 
radio morphology, 
showing that BL Lac objects have similar extended radio properties to FR Is \citep[e.g.,][]{anto85} but, 
however, there are some indications that the radio properties of at least some BL Lac objects are similar 
to FR IIs rather than FR Is \citep[e.g.,][]{koll92,stan97}. There is also evidence that some BL Lac objects 
share the luminosity properties of FR IIs \citep{cass99}. 
The average $B$-band luminosity of low redshift FR I and FR II RGs is $M_{FRI}=-22.1\pm0.6$ and 
$M_{FRII}=-21.9\pm0.6$, respectively, obtained by \citet{govo00}. 
BL Lac hosts appear, therefore, to be slightly fainter than both FR Is and FR IIs, but based on 
Kolmogorov-Smirnov statistics, the luminosity distributions for the BL Lac hosts and 
the combined sample of RGs are indistinguishable. 
Our result suggests that both FR I and FR II RGs can be considered as a parent population of BL Lac objects, 
consistent with the result obtained by \citet{falo99}.

The average effective radius of the BL Lac host galaxies in the $B$-band is $R_e=7.6\pm3.2$ kpc, 
consistent with results from previous studies \citep[e.g.,][]{nils03,falo99} that BL Lac objects are hosted in 
large elliptical galaxies. BL Lac hosts are of similar size in the $B$-band to quasar host galaxies 
($R_e$ = 6.0$\pm2.7$ kpc; \citet{jahn04}) 
but they are slightly smaller than RGs in $R$-band ($R_e=13.6\pm8.7$ kpc); \citet{govo00}). Note, however, that the latter comparison may be biased by a luminosity-dependent selection effect.



It is well known that the effective radius of inactive late-type galaxies 
increases toward shorter wavelengths \citep[e.g.,][]{moll01}. 
This dependence delineates the morphology of SF regions in the galaxies, in the sense that 
SF (blue bands) occurs in the spiral arms while the old stellar population (red bands) 
dominates the bulge component. 
On the other hand, \citet{degr98} found no change in the effective radius from $B$- to $K$-band 
for inactive early-type galaxies (up to Sa spirals). 
The average effective radii of our BL Lac host galaxy sample 
are $R_{e}(U)=6.4\pm2.3$ kpc, $R_{e}(B)=7.6\pm3.4$ kpc and $R_{e}(V)=7.4\pm2.6$ kpc, 
comparable to those in the $H$-band ($R_{e}(H)=7.8\pm4.3$ kpc; \citep{koti04}) and in the $R$-band 
($R_{e}(R)=9\pm5$ kpc; \citep{falo99}). 
In agreement with the result for inactive ellipticals \citet{degr98}, we find that the effective radius of 
the early-type AGN host galaxies does not decrease with wavelength. If anything, one can note a slight 
increasing trend. However, note that this result should be taken with caution because of 
the relatively large errors in determining the effective radius depending on the well known 
degeneracy between surface brightness and effective radius.




\subsection{Kormendy relation}

According to the Kormendy relation, there is a tight relation between the effective radius $R_e$ 
and the surface brightness $\mu_e$ \citep{korm77,korm89}. This relation is a 2D projection of 
the 3D Fundamental Plane \citep[e.g.,][]{dres87,djor87} 
that links $R_e$ and $\mu_e$ with the stellar velocity dispersion $\sigma$. 
This relation, well established for inactive elliptical galaxies and the bulges of spiral galaxies 
in nearby clusters \citep[e.g.,][]{jorg93,jorg96}, is related to the morphology and the dynamical structure of 
the galaxies and gives important information about their formation processes, indicating underlying 
regularity within the galaxy populations.

\begin{figure}
\centering
\includegraphics[width=9cm]{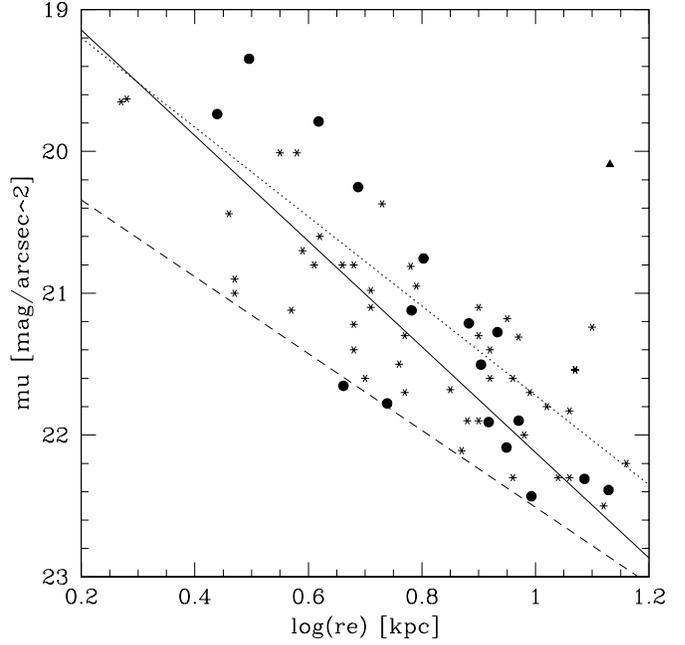}
\caption{The $B$-band $\mu_{e}$$ - R_e$ Kormendy relation for the BL Lac 
host galaxies (filled symbols), and for inactive elliptical galaxies (asterisks; \citet{jorg93} and \citet{reda05}). 
The solid, dotted and long dashed lines represent the best linear fit 
for BL Lac hosts, inactive ellipticals (\citet{jorg93} and \citet{reda05}), and low redshift RGs (\citet{fasa96}), respectively. 
\label{korm}}
\end{figure}

Fig.~\ref{korm} presents the $B$-band Kormendy relation for our sample of 18 BL Lac objects, compared 
with that for nearby RGs \citep{govo00}, isolated inactive 
elliptical galaxies \citep{reda05} and inactive ellipticals in the Coma cluster \citep{jorg93}. 
The corresponding surface brightnesses $\mu_{e}$ and effective radii $R_e$ for the BL Lac hosts are given 
in Table~\ref{results}. 
The surface brightnesses of the host galaxies were corrected for Galactic extinction and cosmological 
dimming ($10 \times$ log(1+z)). 
Note that the surface brightnesses $\mu_{e}$ and effective radii $R_e$ refer to the isophote that contains 
half of the total luminosity of the galaxy. Another commonly used definition for $\mu_{e}$ and $r_e$ is to 
derive these values directly from the de Vaucouleurs fitting of the galaxy, but that definition is model dependent.

The best-fit linear relation for the BL Lac hosts is $\mu$$_e$ = 3.3 log R$_e$ (kpc) + 18.4 mag arcsec$^{-2}$. 
It is somewhat steeper than that for inactive elliptical galaxies and RGs. It is, however, consistent 
with the relation obtained by \citet{falo99} and \citet{koti04} for BL Lac hosts in the $R$- and $H$-bands, 
indicating that the dynamical structure of the hosts does not change with wavelength. 
Note that one BL Lac object (MS 1552.1+2020, filled triangle in Fig.~\ref{korm}) clearly deviates from 
the $\mu_{e}-R_{e}$-relation of the sample, 
towards large effective radius and high surface brightness. Note that MS 1552.1+2020 also has one of 
the bluest host galaxies in the sample ($B$--$H$=3.2; Table~\ref{colours}), and its displacement from 
the Kormendy relation may indicate recent SF as suggested for similar cases among 
inactive ellipticals by \citet{reda05}. Overall, it seems that BL Lac 
hosts are not very far from the Fundamental Plane of inactive elliptical galaxies, indicating that they are 
dynamically similar. It supports the view that BL Lac hosts 
and normal ellipticals have a similar formation process and that all elliptical 
galaxies may have experienced a phase of nuclear activity in the past with 
little influence on the global structure of the galaxy.

\subsection{Host galaxy colours}

The host galaxies of all the BL Lac objects in our sample were previously studied 
in the optical $R$-band and in the NIR $H$-band \citep{falo99,koti04}. 
The new observations presented here represents the first host galaxy study performed in the 
blue part of the spectrum. 
The combination of blue, optical and NIR data allows us to assess the issue of the optical-NIR colours 
of the BL Lac host galaxies. While the $H$-band is sensitive to old stellar populations, the blue domain of 
the spectrum is especially important in studying SF and the presence of a young stellar population component in the galaxies.

The integrated rest-frame $U$--$B$, $B$--$V$ and $B$--$R$ colours of our BL Lac sample are given in 
Table~\ref{colours}. The average colours of the BL Lac hosts, 
compared to those observed in RLQ and RQQ host galaxies \citep{jahn04}, RGs \citep{govo00}, 
and inactive elliptical galaxies \citep{pele90,bowe92a,colb01}, and theoretically 
predicted for inactive ellipticals \citep{fuku95,fioc99} are presented in Table~\ref{col_average}.
Because $U$-band data is available only for eight objects in our sample, 
we prefer to use the $B$--$H$ colour as the longest baseline colour.

\begin{figure}
\centering
\includegraphics[width=18cm]{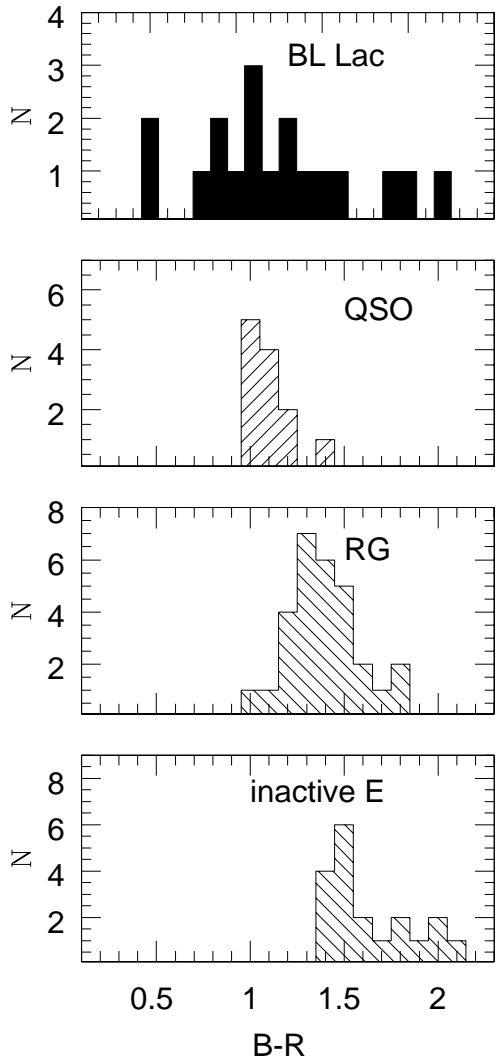}
\caption{The distribution of the $B$--$R$ colour for BL Lac hosts (top panel; this work), 
low redshift ($z<0.2$) RLQ and RQQ hosts (second panel; \citet{jahn04}), low redshift RGs 
(third panel; \citet{govo00}) and inactive elliptical galaxies 
(bottom panel; \citet{colb01} and \citet{pele90}).
\label{br_histogram}}
\end{figure}

The average $B$--$H$ colour 
of the BL Lac hosts is $B$--$H$=$3.5\pm0.5$ that is slightly redder than that of RLQ and RQQ hosts 
($B$--$H$=$2.9\pm0.3$, \citet{jahn04}) but slightly bluer than that of inactive elliptical galaxies 
($B$--$H$=$3.8\pm0.3$, \citet{colb01}). 
The same situation applies when considering the $B$--$V$ and $B$--$R$ colours (Table~\ref{col_average}).
In fact, the optical colours of the BL Lac hosts are very 
similar to those of intermediate/late-type (Sb - Sbc) inactive galaxies
with significant ongoing SF (Table~\ref{col_average}). Fig.~\ref{br_histogram} shows the 
distribution of the $B$--$R$ colour for BL Lac hosts, quasar hosts \citep{jahn04}, RGs \citep{govo00} and 
inactive ellipticals \citep{colb01}. BL Lac hosts clearly exhibit a much wider colour distribution and have 
a significantly bluer host galaxy population that is not present for inactive ellipticals.

\setlength{\tabcolsep}{0.8mm}
\begin{table*}
\centering
\caption{Colours and colour gradients of the BL Lac host galaxies.$^{\mathrm{a}}$}
\label{colours}
\begin{tabular}{lccccccc}
\hline
\noalign{\smallskip}
Name & $U$--$B$ & $B$--$V$ & $B$--$R$ & $R$--$H$ & $\Delta(U$--$B)/\Delta(log r)$ &$\Delta(B$--$R)/\Delta(log r)$& $\Delta(B$--$H)/\Delta(log r)$\\
& & & & & & & \\
(1) & (2) & (3) & (4) & (5) & (6) & (7) & (8)\\
\noalign{\smallskip}
\hline
\noalign{\smallskip}
1ES 0229+200   &  0.4 & 0.3 & 2.0 & 2.5 & 0.92   & --0.44 &--0.17\\
PKS 0521-365   &      & 0.9 & 0.9 & 2.3 &        & --1.02 &--1.66\\
PKS 0548-322   &      &     & 1.1 & 2.1 &        &   1.10 &--0.90\\
1H 0706+591    &  0.9 & 0.8 & 1.3 & 2.2 & --0.01 &   0.17 &--0.12\\
MRK 421        & --0.2 &     & 2.2 & 2.2 & 0.17  & --0.91 &--0.31 \\
MRK 180        &  0.4 &     & 0.5 & 2.7 & 0.75   & --1.24 &--0.91\\
1ES 1212+078   &      &     & 1.9 & 2.2 &        &   0.48 &--0.85\\
MS 1229.2+6430 &      &     & 1.3 & 0.9 &        & --0.12 &--0.74\\
1ES 1255+244   &      &     & 0.9 & 2.1 &        & --1.08 &--0.91\\
PG 1418+546    &      &     & 1.4 & 2.2 &        & --0.21 & 0.86\\
1ES 1426+428   &      &     & 0.8 & 2.5 &        &   0.90 & 1.19\\
1ES 1440+122   &      &     & 1.6 & 2.4 &        &   0.16 &--0.11\\
AP LIBRAE      & --0.5 &     & 1.1 & 2.2 & 0.17  &   0.43 &--0.31\\
MS 1552.1+2020 &      &     & 0.5 & 2.7 &        & --0.41 &--0.37\\
MRK 501        &  0.4 & 1.2 & 1.1 & 2.4 & 0.01   &   0.64 &--0.62\\
I Zw 187       &  0.5 & 1.1 & 1.2 & 2.3 & 0.63   & --0.32 &--0.54\\
3C 371         &  0.2 & 0.8 & 1.5 & 1.9 & --0.44 & --1.26 &--1.51\\
BL LACERTAE    &      &     & 1.0 & 2.4 &        &   0.63 &--1.70\\
\noalign{\smallskip}
\hline
\end{tabular}
\begin{list}{}{}
\item[$^{\mathrm{a}}$]
Column (1) gives the name of the object; (2) the $U$--$B$ colour; 
(3) the $B$--$V$ colour; (4) the $B$--$R$ colour; (5) the $R$--$H$ colour 
(from \citet{koti04}); (6) the $U$--$B$ colour gradient; 
(7) the $B$--$R$ colour gradient and (8) the $B$--$H$ colour gradient of the host galaxy.
\end{list}
\end{table*}

\setlength{\tabcolsep}{0.8mm}
\begin{table*}
\centering
\caption{Average colours of the host galaxies.$^{\mathrm{a}}$}
\label{col_average}
\begin{tabular}{lccccc}
\hline
\noalign{\smallskip}
The sample & $U$--$B$ & $B$--$V$ & $B$--$R$ & $R$--$H$ & $B$--$H$ \\
& & & & &\\
(1) & (2) & (3) & (4) & (5) & (6)\\
\noalign{\smallskip}
\hline
\noalign{\smallskip}
BL Lacs (this work)        & $0.3\pm0.4$ & $0.8\pm0.2$ & $1.2\pm0.5$ & $2.2\pm0.4$ & $3.5\pm0.5$ \\
Quasars \citep{jahn04}     &             & $0.5\pm0.1$ & $1.1\pm0.1$ & $1.9\pm0.2$ & $2.9\pm0.3$ \\
RGs \citep{govo00}         &             &             & $1.4\pm0.2$ &             &             \\
Ellipticals \citep{pele90} & $0.6\pm0.1$ &             & $1.6\pm0.1$ &             &             \\
Ellipticals \citep{bowe92a}&             &             &             &             & $3.8\pm0.2$ \\
Ellipticals \citep{colb01} &             &              & $1.5\pm0.2$ &             &             \\
Ellipticals \citep{fuku95} &             &  1.0        & 1.6         &             &             \\
Ellipticals \citep{fioc99} &             &             &             & 2.2         & 3.8         \\
Spirals (Sb) \citep{fuku95,fioc99} &       &  0.7        & 1.2         & 2.1         & 3.3            \\

\noalign{\smallskip}
\hline
\end{tabular}
\begin{list}{}{}
\item[$^{\mathrm{a}}$]
Column (1) gives the sample; and columns (2) -- (6) the average $U$--$B$, $B$--$V$, $B$--$R$, $R$--$H$ 
and $B$--$H$ colour, respectively. 
\end{list}
\end{table*}
\setlength{\tabcolsep}{1.5mm}

It has long been known that 
the integrated colours of elliptical galaxies become redder toward higher luminosity (mass). 
\citep[e.g.,][]{bowe92a,koda97}. 
This colour-magnitude relation links the properties of the stellar populations of early-type galaxies 
with their structural properties and provides important information about their formation and evolution. 
It depends on the combined effects of age and metallicity on the dominant stellar population 
(more massive galaxies are both older and more metal-rich than less massive galaxies). 
It is identical in different galaxy clusters, such as in Virgo and Coma clusters \citep{bowe92a} 
and overall shows little dependence on environment \citep{terl01,bern03} and redshift 
\citep[e.g.,][]{arag93,koda98,hold04}, indicating that massive ellipticals 
formed in an intense starburst at high redshift followed by passive evolution, with no major SF episodes 
since z $\sim$2.

\begin{figure}
\centering
\includegraphics[width=17cm]{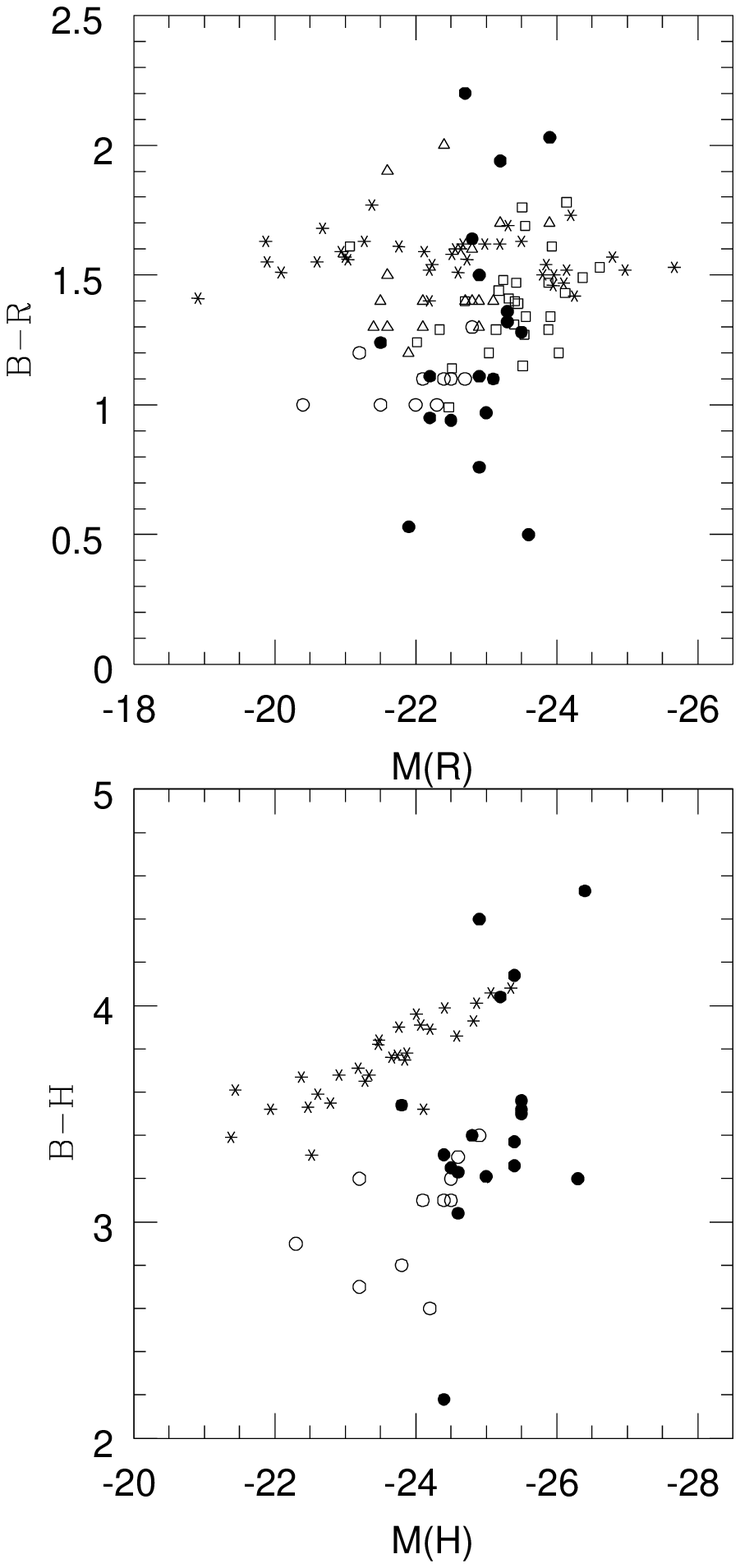}
\caption[]{Upper: The $B$--$R$ vs. $R$ colour-magnitude diagram for BL Lac host galaxies (filled circles), RLQ and RQQ hosts (open circles; \citet{jahn04}), RGs (open squares; \citet{govo00}) and inactive early-type galaxies (asterisks; \citet{pele90,colb01}). Lower: The $B$--$H$ vs. $H$ colour-magnitude diagram for BL Lac host galaxies and comparison samples. Open 
circles are from \citet{jahn04} and asterisks from \citet{bowe92a}.
\label{brbh}}
\end{figure}

The $B$--$R$ vs. $R$ and $B$--$H$ vs. $H$ colour-magnitude diagrams for the BL Lac hosts (this work), 
low redshift RLQ and RQQ hosts \citep{jahn04}, RGs \citep{govo00}, inactive ellipticals \citep{pele90,colb01} 
and ellipticals in Virgo and Coma clusters \citep{bowe92a} are 
presented in Fig.~\ref{brbh}. 
It is evident that the BL Lac hosts do not follow the relatively tight 
colour-magnitude relation of inactive ellipticals. 
Instead, they have a significantly broader colour distribution and the majority of them 
appear to be bluer than inactive ellipticals of similar luminosity. Indeed, such colours are more similar 
to those found in intermediate/late-type inactive galaxies that have significant recent SF. 
This result is consistent with the colours of low redshift RGs \citep{govo00} and 
quasar hosts \citep{jahn04}. Although this colour difference may indicate that the 
colour-magnitude relation for elliptical galaxies breaks down at high luminosities as suggested by 
\citet{govo00} and \citet{koti04}, note that the elliptical galaxies cover the same luminosity range as the BL Lac hosts. 
A more likely explanation 
for the blue colours of BL Lac hosts (and the other AGN hosts in the diagrams) 
is that they have experienced recent SF. 
The wide colour distribution indicates a range of timescale since the latest SF episode, 
such that the bluest hosts have experienced the most recent SF whereas the reddest hosts have 
experienced little or no SF and are dominated 
by an old stellar population similar to those in inactive ellipticals. 
Especially note that there is one BL Lac host exhibiting red $B$--$R$ colour ($B$--$R>2.0$) and two hosts having red $B$--$H$ colour ($B$--$H>4.4)$, 
redder than any of the inactive ellipticals. 
From this point of view, the bluest hosts may hold evidence from an 
event that triggered both the nuclear activity and the strong SF. This is consistent with 
the blue colours ($B$--$R=1.1\pm0.1$) in RLQ and RQQ hosts \citep{jahn04}. Similarly blue host colours 
were also found in low redshift quasars by \citet{scha00}.

\subsection{Colour gradients of the host galaxies}

It is well known that nearby inactive elliptical galaxies do not have uniform colour but instead have 
negative colour gradients, i.e. they become bluer with increasing radius \citep[e.g.,][]{pele90}. 
These colour gradients have been widely interpreted as due to radial variations in the dust content 
and/or SF history of the galaxies \citep[e.g.,][]{kim89,pele90,goud94,mich00}.
As dust reddening is relatively small in ellipticals, while normal early-type galaxies can be dusty 
\citep{tran01}, most of the colour gradients are usually 
ascribed to the combined effect of stellar metallicity and age gradients across the galaxies, 
with the outer regions being younger and/or having lower metallicity.

\begin{figure*}
\centering
\includegraphics{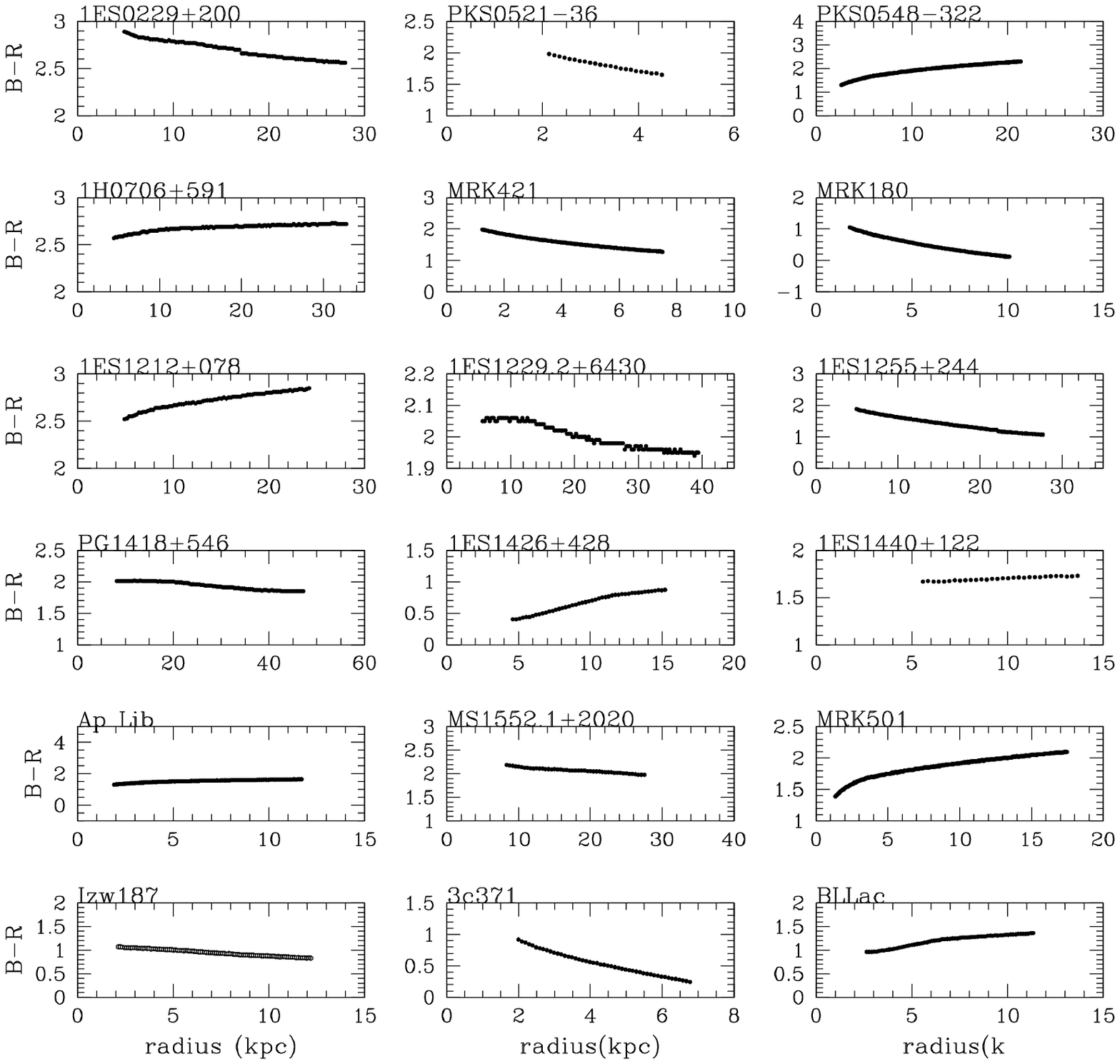}
\caption{The $B$-$R$ colour profiles for the BL Lac host galaxies, derived using the $R$-band data 
from \citet{falo99,scar00b} and \citet{koti04}.}
\label{grad_br}
\end{figure*}

For the BL Lac host galaxies, we derived the radial $U$--$B$, $B$--$R$ and $B$--$H$ colour gradients using 
the $R$- and $H$-band host galaxy luminosity profiles from \citet{koti98b,falo99,koti04,scar00b}. 
The $B$--$R$ colour profiles are presented in Fig.~\ref{grad_br} and all colour gradients are 
presented in Table~\ref{colours}. 
Each gradient was cut at the radius $<$ 2 arcsec to avoid any contamination of the host galaxy 
from the bright nucleus. 
The colour as a function of logarithmic radius is usually well represented by a linear fit. 
As expected for early-type galaxies, most of the BL Lac host galaxies exhibit a negative colour gradient 
with average 
$\Delta(B$--$R)/\Delta(logr)=-0.14\pm0.75$. The same negative trend was also found for the $R$--$H$ colour 
gradients of BL Lacs $(\Delta(R$--$H)/\Delta(log r)=-0.38\pm0.60$ by \citet{koti04}). 
The amplitude of the $B$--$R$ colour gradients of the BL Lac hosts is consistent with 
the average $B$--$R$ gradient of RGs $\Delta(B$--$R)/\Delta(log r)=-0.16\pm0.17$ \citep{govo00} and inactive 
ellipticals $\Delta(B$--$R)/\Delta(log r)=-0.09\pm0.07$ \citep{pele90}, 
but the distribution is much wider than that of the RGs and inactive ellipticals. Similar distribution was 
also obtained for the $R$--$H$ gradient of BL Lac hosts by \citet{koti04}. 

However, note that some BL Lac host galaxies have little evidence for colour variation or even have an 
inverted (positive) colour gradient, the steepest of them in PKS 0548-322, 1ES 1426+428 and MRK 501. 
Similarly steep positive $R$--$H$ colour gradient for 1ES1426+428 was previously observed by 
\citet{koti04}. Since the $B$--$H$ colour is more sensitive to the dust content than the $R$--$H$ colour, 
it indicates that the positive $B$--$R$ and $R$--$H$ gradients of this target are due to radial variations 
in the dust content of its host galaxy. 
Indeed, recent HST observations of RGs \citep[e.g.,][]{mart00,trem07} 
have shown that there is a significant number of RGs that contain dust in a variety of spatial distributions, 
such as circumnuclear disks and dust lanes at kpc scales. On the other hand, dust is not a reasonable 
explanation for e.g. MRK 501 which has a positive $B$--$R$ but a negative $R$--$H$ gradient. In this case,  
the inverted profiles indicates SF in the inner region of the host galaxy because the host galaxy is 
well resolved. That indication is also supported by recent spectroscopic observations of ongoing SF in 
the nuclear regions of a BL Lac object (PKS 2005-489; \citet{bres06}).

\begin{figure*}
\centering
\includegraphics{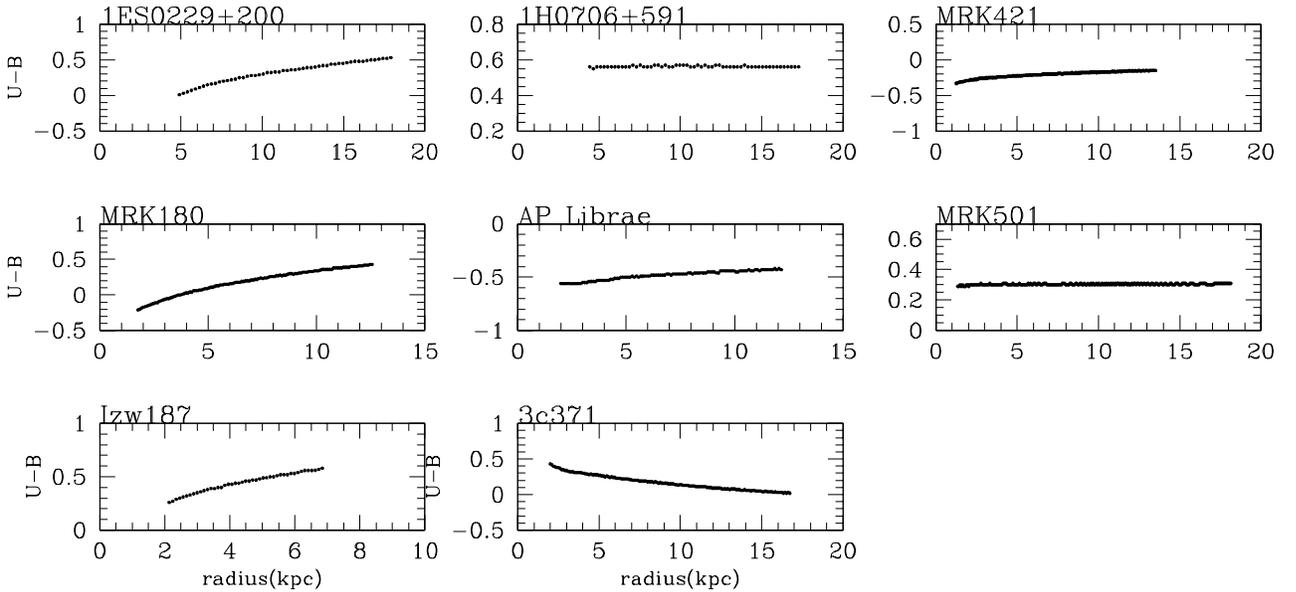}
\caption{The $U$--$B$ colour profiles for BL Lac hosts.}
\label{grad_ub}
\end{figure*}

Fig.~\ref{grad_ub} shows the $U$--$B$ colour profiles for the eight BL Lac host galaxies for 
which we have the available data (Table~\ref{colours}). 
All the host galaxies, except one (3C 371), have a positive colour gradient. 
3C 371 has some signature of recent interaction \citep{nils97} which might be the reason for the negative $U$--$B$ colour gradient. 
Average colour gradient for all eight objects is $\Delta(U$--$B)/\Delta(log r)=0.28\pm0.46$, 
although for four objects the gradient is very flat.

\subsection{Stellar population model fits}

The $UBV$ colours combined with the previously observed $R$- and $H$-band data can be 
used in conjunction with stellar synthesis population models to estimate the approximate ages of the 
recent SF episodes in the BL Lac host galaxies. 
In this context, the $U$- and $B$-bands are especially important as they provide photometry 
shortward of the 4000 \AA ~break, without which only approximate average ages can be determined. 
Note also that dust reddening has little effect, since any reddening only would imply 
even bluer intrinsic colours. The SED modelling utilizes 
all the colour information simultaneously by fitting the SED of evolution synthesis model to 
calibrated fluxes of the host galaxies at different wavelengths. Each wavelength band 
represents one data point in the SED of the object. Comparing fits made for different 
model spectra gives an estimate of the ages of the dominant stellar population components in 
the host galaxy. Such SED modelling has been performed by \citet{jahn04} for low redshift 
RLQ and RQQ hosts. For the analysis we used the PEGASE2 evolutionary 
model \citep{fioc97} and a single stellar population (SSP) model of single metallicity 
because the age-metallicity degeneracy in models cannot be resolved with multicolour data only. 
We used instantanous burst models which assume that the young stellar population is 
formed in a short burst with an IMF and evolves passively thereafter. 


Synthesized galaxy spectra composed from libraries of stellar spectra can be used 
to derive the colours of the galaxies over a large wavelength range from UV to NIR for 
different ages and metallicities. For generating the synthetic 
spectra we used the \citet{scal86} initial mass function, solar metallicity and 
five different model ages (0.1, 0.7, 2, 6 and 14 Gyr), to allow a direct comparison with 
the results of \citet{jahn04}.

From the synthesized spectra we have derived the $U$--$B$, $B$--$V$, $B$--$R$, $R$--$H$ and $B$--$H$ colours 
for each model ages and compared those with the observed colours of each BL Lac using an iterative 
least squares fitting, with age as the only free parameter.
The results of the fitting are presented in Table~\ref{colour_fit}, where the best fitting model is marked as 1 
and the poorest fitting model as 3.



\begin{table}
\centering
\caption{Colours of the single SSP fit compared to the colours of the BL Lac sample. 
1 represents the best and 3 the worst fit.}
\label{colour_fit}
\begin{tabular}{llllll}
\hline
\noalign{\smallskip}
Name & 0.1 & 0.7 & 2.0 & 6.0 & 14.0 \\
& Gyr & Gyr & Gyr & Gyr & Gyr \\
\noalign{\smallskip}
\hline
\noalign{\smallskip}
1ES 0229+200   &     &    & 3   & 1   & 2    \\
PKS 0521-365   & 3    & 1   & 2   &    &     \\
PKS 0548-322   & 2    & 1   & 3   &    &    \\
1H 0706+591    &     & 2   & 1   & 3   &     \\
MRK 421        &     &    & 2   & 1   & 3    \\
MRK 180        & 3    & 1   & 2   &    &     \\
1ES 1212+078   &     &    & 1   & 2   & 3    \\
MS 1229.2+6430 & 2    & 1   & 3   &    &     \\
1ES 1255+244   & 2    & 1   & 3   &    &     \\
PG 1418+546    &     & 2   & 1   & 3   &     \\
1ES 1426+428   & 3    & 2   & 1   &    &     \\
1ES 1440+122   &     &    & 1   & 2   & 3    \\
AP LIBRAE      & 3    & 1   & 2   &    &     \\
MS 1552.1+2020 & 1    & 2   & 3   &   &     \\
MRK 501        &     & 2   & 1   & 3   &     \\
I Zw 187       &     & 2   & 1   & 3   &     \\
3c371          &     & 2   & 1   & 3   &     \\
BL LACERTAE    & 3    & 1   & 2   &    &     \\
\noalign{\smallskip}
\hline
\end{tabular}
\end{table}

The fitting results indicate that the best fitting models are generally consistent with 
a young/intermediate age stellar population. Only for two BL Lacs (1ES 0229+200 and MRK 421), 
the best fitting model is older than 2 Gyr and is best consistent with the 6 Gyr model. 
Note that these objects also have the reddest host galaxies in the sample ($B$--$H\sim4.4$, 
see Table~\ref{colours}). The model fitting is thus consistent with the host galaxy colours, 
suggesting that the host galaxies of these objects are dominated by an old stellar population, with no 
evidence for a young population. 
Eight objects are best fit with the 2 Gyr model, while the remaining eight objects in the sample require 
an even younger model, 0.7 Gyr, and in the case of MS 1552.1+2020, the 0.1 Gyr model. 
Note that MS 1552.1+2020 deviates from the sample also in the Kormendy relation (see above). 

These results are in excellent agreement with similarly young stellar populations 
found by \citet{jahn04} for RLQ and RQQ host galaxies obtained from population synthesis model. 
They found the 2 Gyr model to best fit their data in 12 out of their 19 objects, and an even 
younger population (0.7 Gyr) in five objects, while an older population 
(6 Gyr) was preferred in only two objects. 
Note that, as is the case for the quasar hosts \citep{jahn04}, none of the elliptical BL Lac hosts 
are best modeled with the 14 Gyr model, i.e. a very old, evolved population as would be expected for 
early-type host galaxies. On the other hand, for both types of AGN hosts, there is also little evidence for 
massive ongoing starbursts with a significant very young population (age $<<$ 1 Gyr). 
Together these results are in good agreement with the colour information (section 3.3) and strongly support 
the idea that the blue colours of the early-type host galaxies of moderately luminous AGN 
are caused by them having experienced a relatively recent SF episode. 
Further support for this conclusion is provided by spectroscopic evidence for young/intermediate age populations 
in AGN hosts \citep[e.g.,][]{nola01,raim05}. On the other hand, both the predominantly late-type hosts of 
low luminosity AGN \citep[e.g.,][]{koti94,scha00,jahn04} and the elliptical hosts of very luminous AGN 
\citep[e.g.,][]{dunl03} exhibit similar colours to their inactive counterparts.

The blue colours and steep colour gradients of early-type AGN hosts found in an increasing number of studies
are most likely caused by a young stellar population, 
and indicate a link between SF and the onset of the nuclear activity, both likely triggered by a tidal 
interaction or a minor or major merging event. However, the lack of obvious signs of interaction 
(companion galaxies, tidal tails, disturbed morphology) in the close environment of the large majority of 
the host galaxies may require 
a significant time delay (at least hundreds of Myr) between the event with associated SF episodes and 
the start of the nuclear activity. Such a time delay is indeed 
predicted by simulations of gas reaching the galaxy center after a merger/interaction \citep[e.g.,][]{lin88}.

\section{Conclusions}

We have presented $B$-band imaging of a sample of 18 low redshift ($z<0.3$) BL Lac objects 
for which their host galaxies were previously resolved in the optical $R$- and NIR $H$-bands. 
Subsamples consisting of nine and five BL Lacs were also imaged in the $U$- and $V$-bands, respectively. 
All the BL Lac objects were clearly resolved in the $B$- and $V$-bands while 8/9 objects were resolved 
in the $U$-band. All the host galaxies are well described by an elliptical de Vaucouleurs model. 
These data are combined with previous optical and NIR data to study the optical-NIR colours and colour gradients 
of the host galaxies.

BL Lac host galaxies are luminous (massive) 
and large elliptical galaxies having average $B$-band absolute magnitude 
$M_{B}=-21.6\pm0.7$ and average effective radius $R_{e}=7.6\pm3.2$ kpc. They are indistinguishable from 
low redshift radio galaxies but brighter than inactive elliptical galaxies. The $B$-band Kormendy relation 
for BL Lac hosts is slightly steeper than that of RGs but does not deviate significantly from the Fundamental Plane 
of inactive elliptical galaxies. This indicates that BL Lac hosts are dynamically similar to normal ellipticals 
and that the active nuclear phase does not have any significant influence on the dynamical structure 
of the galaxy. Thus it is possible that all massive galaxies can experience an AGN phase.

The distributions of the integrated blue/NIR colours ($B$--$H=3.5\pm0.5$; $B$--$R=1.2\pm0.5$) and colour gradients 
$(\Delta(B$--$R)/\Delta(log r)=-0.14\pm0.75)$ of the BL Lac hosts 
are much wider than those for normal ellipticals with old stellar populations, and many BL Lac objects have 
bluer hosts and/or steeper colour gradients than those in normal ellipticals. 
The blue colours and steep colour gradients are most likely caused by a young stellar population, 
and indicate a link between SF caused by an interaction/merging event and the onset of the nuclear activity. 
Many targets have inverted (positive) colour gradients which in some cases, e.g. 1ES 1426+428, 
are most likely caused by radial variations in the dust content of the host galaxy, 
but in some other cases may suggest star formation in the inner regions of the host galaxy. 
This result is corroborated by stellar population modelling, indicating young/intermediate age 
populations in the majority of the sample, in agreement with low redshift quasar hosts. 
However, the lack of obvious signs of interaction may require 
a significant time delay between the event with associated SF episodes and 
the start of the nuclear activity.

Future work in this area should address the correlation between colour and statistics of companion galaxies and 
morphological disturbances as an indicator for the interaction, for well defined samples of AGN hosts and 
inactive galaxies. In a forthcoming paper, we shall present NIR spectroscopy of 
BL Lac hosts and RGs with blue colours, to analyze their stellar content and SF properties, 
based on emission and absorption line diagnostics, in more detail than is affordable with imaging. 

\begin{acknowledgements}
Based on observations made with the Nordic Optical Telescope, operated
on the island of La Palma jointly by Denmark, Finland, Iceland,
Norway, and Sweden, in the Spanish Observatorio del Roque de los 
Muchachos of the Instituto de Astrofisica de Canarias. 
This work was supported by the Italian Ministry for University and Research (MIUR) 
under COFIN 2002/27145, ASI-IR 115 and ASI-IR 35, ASI-IR 73 
and by the Academy of Finland (projects 8201017 and 8107775). 
This research has made use of the NASA/IPAC Extragalactic Database {\em(NED)} which is operated by 
the Jet Propulsion Laboratory, California Institute of Technology, 
under contract with the National Aeronautics and Space Administration. 
We thank Valentin Ivanov for obtaining for us the observations with NTT/EMMI. 

\end{acknowledgements}
\begin{figure*}
\centering
\includegraphics{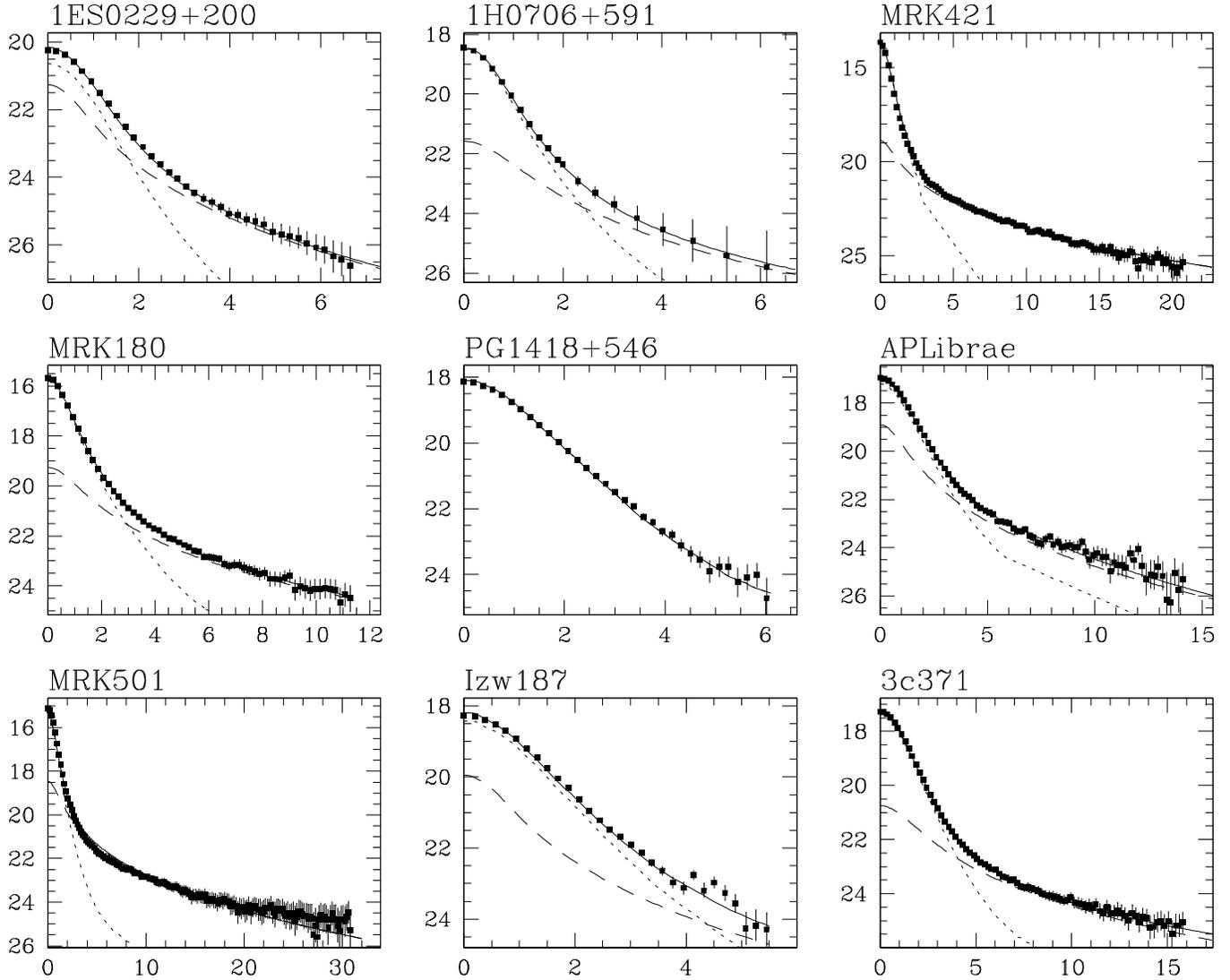}
\caption{The observed $U$-band radial luminosity profiles (solid points with error bars) 
for each BL Lac, overlaid with the PSF model (dotted line), the de Vaucouleurs $r^{1/4}$ model 
(long-dashed line) and the fitted PSF+host galaxy model profile (solid line). The X-axis is in 
arcsec and the Y-axis in mag arcsec${^-2}$.}
\label{profileu}
\end{figure*}

\begin{figure*}
\centering
\includegraphics{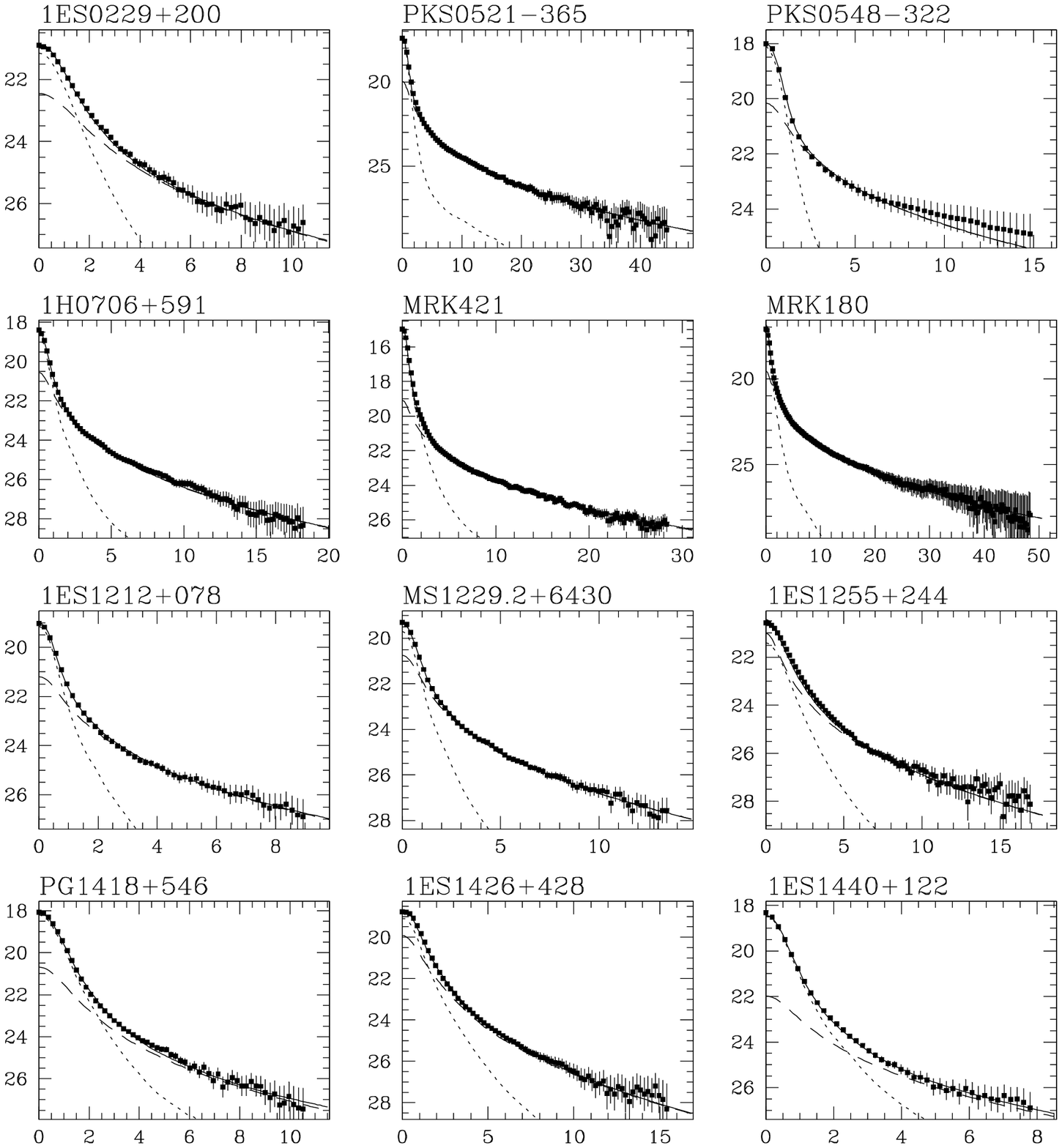}
\caption{The observed $B$-band radial luminosity profiles and model profiles. For explanation, see the caption of 
Fig.~\ref{profileu}.}
\label{profileb}
\end{figure*}
\addtocounter{figure}{-1}

\begin{figure*}
\centering
\includegraphics{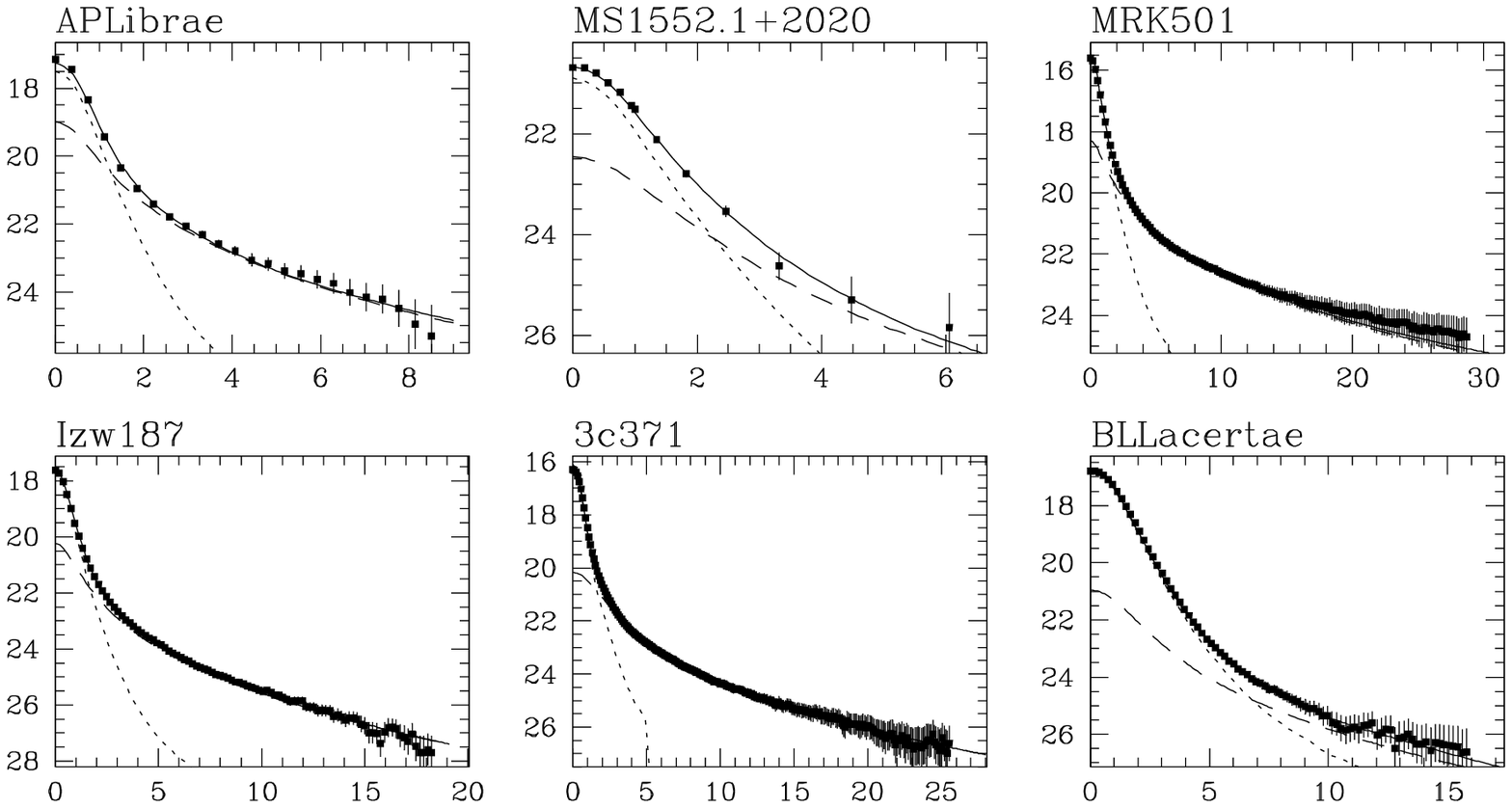}
\caption{(continued)}
\end{figure*}

\begin{figure*}
\centering
\includegraphics{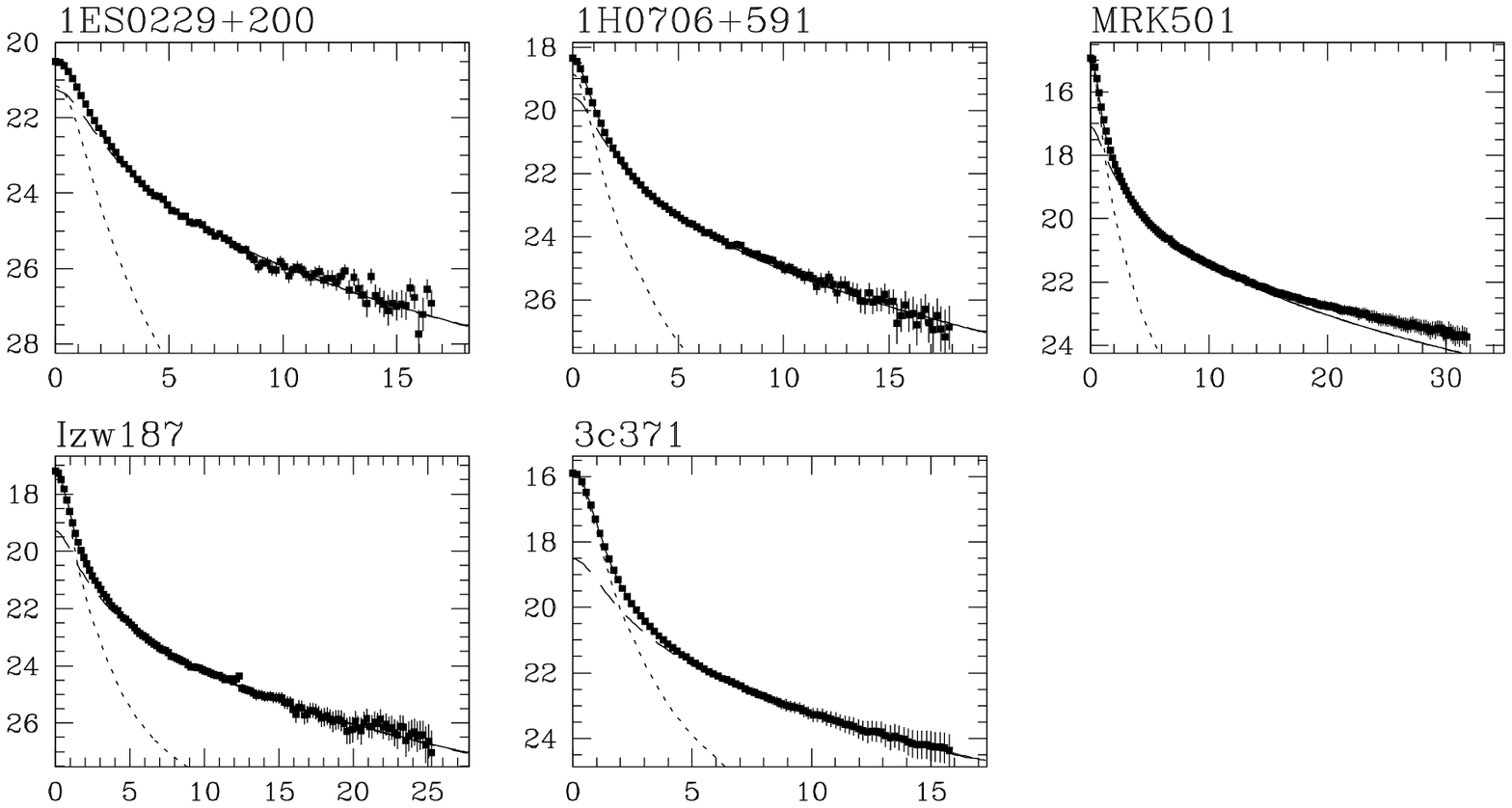}
\caption{The observed $V$-band radial luminosity profiles and model profiles. For explanation, see the caption of 
Fig.~\ref{profileu}.}
\label{profilev}
\end{figure*}

\bibliographystyle{aa} 
\bibliography{totahy.bib}

\end{document}